# Viewpoint

## Finally, results from Gravity Probe B

**Clifford M. Will**
*McDonnell Center for the Space Sciences and Department of Physics, Washington University, St. Louis, One Brookings Drive, St. Louis, MO 63130, USA*

Published May 31, 2011

*Nearly fifty years after its inception, the Gravity Probe B satellite mission delivers the first measurements of how a spinning gyroscope precesses in the gravitational warping of spacetime.*

Subject Areas: **Astrophysics, Gravitation**

The great blues singer Etta James' signature song begins, "At laaasst, my love has come along . . . ." This may have been the feeling on May 4th when NASA announced the long-awaited results of Gravity Probe B [1], which are appearing now in *Physical Review Letters*[2]. Over 47 years and 750 million dollars in the making, Gravity Probe B was an orbiting physics experiment, designed to test two fundamental predictions of Einstein's general relativity.

According to Einstein's theory, space and time are not the immutable, rigid structures of Newton's universe, but are united as spacetime, and together they are malleable, almost rubbery. A massive body warps spacetime, the way a bowling ball warps the surface of a trampoline. A rotating body drags spacetime a tiny bit around with it, the way a mixer blade drags a thick batter around.

The spinning Earth does both of these things and this is what the four gyroscopes aboard the earth-orbiting satellite Gravity Probe B measured. The satellite follows a polar orbit with an altitude of 640 kilometers above the earth's surface (Fig. 1, top). The warping of spacetime exerts a torque on the gyroscope so that its axis slowly precesses—by about 6.6 arcseconds (or 1.8 thousandths of a degree) per year—in the plane of the satellite's orbit. (To picture this precession, or "geodetic effect," imagine a stick moving parallel to its length on a closed path along the curved surface of the Earth, returning to its origin pointing in a slightly different direction than when it started.) The rotation of the Earth also exerts a "frame-dragging" effect on the gyro. In this case, the precession is perpendicular to the orbital plane and advances by 40 milliarcseconds per year. Josef Lense and Hans Thirring first pointed out the existence of the frame-dragging phenomenon in 1918, but it was not until the 1960s that George Pugh in the Defense Department and Leonard Schiff at Stanford independently pursued the idea of measuring it with gyroscopes.

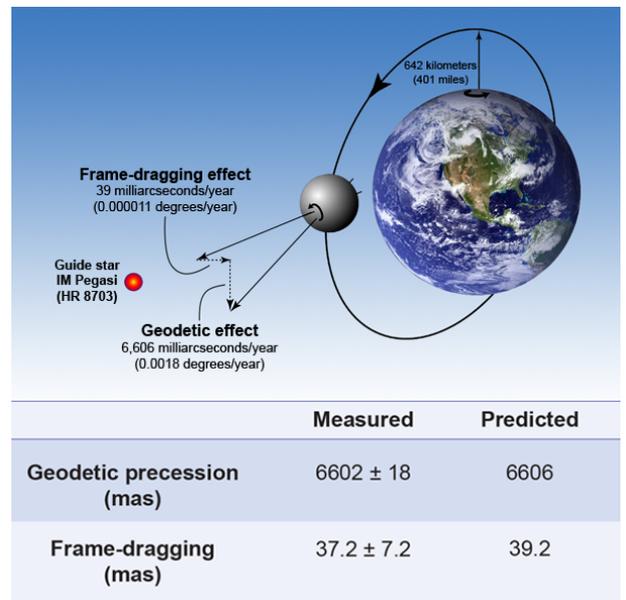

FIG. 1: (Top) Schematic of the orbit of the Gravity Probe B satellite. Two relativistic effects—frame-dragging and the geodetic effect—were expected to cause a precession (at ninety degree angles with respect to one another) of the gyroscopes aboard the satellite. (Bottom) Predicted precessions, given in milliarcseconds (mas), are compared to those measured by GPB. (Credit: C. W. F. Everitt *et al.* [2])

The Gravity Probe B (or GP-B, in NASA parlance) gyroscopes (Fig. 2) are coated with superconducting niobium, such that when they spin, the supercurrents in the niobium produce a magnetic moment parallel to the spin axis. Extremely sensitive magnetometers (superconducting quantum interference detectors, or "SQUIDs") attached to the gyroscope housing are capable of detecting even minute changes in the orientation of the gyros' mag-

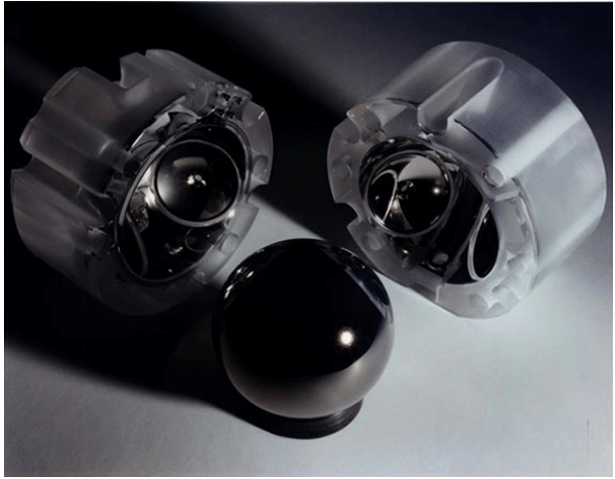

FIG. 2: A gyroscope aboard the Gravity Probe B satellite. Each gyroscope is a fused silica rotor about the size of a ping-pong ball, machined to be spherical and homogeneous to tolerances better than a part per million, and coated with a thin film of superconducting niobium. The gyroscope assembly sat in a dewar of 2440 liters of superfluid helium and was held at 1.8 degrees Kelvin. (Credit: Bob Kahn/Stanford University/Gravity Probe B)

netic moments and hence the precession in their rotation predicted by general relativity.

At the start of the mission, the four gyros were aligned to spin along the symmetry axis of the spacecraft. This was also the optical axis of a telescope directly mounted on the end of the structure housing the rotors. Spacecraft thrusters oriented the telescope to point precisely toward the star IM Pegasi (HR 8703) in our galaxy (except when the Earth intervened, once per orbit). In order to average out numerous unwanted torques on the gyros, the spacecraft rotated about its axis once every 78 seconds.

GP-B started in late 1963 when NASA funded the initial R&D work that identified the new technologies needed to make such a difficult measurement possible. Francis Everitt, a physicist at Stanford and a lead author on the current paper, became Principal Investigator of GP-B in 1981, and the project moved to the mission design phase in 1984. Following a major review of the program by a National Academy of Sciences committee in 1994, GP-B was approved for flight development, and began to collaborate with Lockheed-Martin and Marshall Space Flight Center. The satellite launched on April 20, 2004 for a planned 16-month mission, but another five years of data analysis were needed to tease out the effects of relativity from a background of other disturbances.

Almost every aspect of the spacecraft, its subsystems, and the science instrumentation performed extremely well, some far better than expected. Still, the success of such a complex and delicate experiment boils down to figuring out the sources of error. In particular, having an accurate calibration of the electronic readout from the SQUID magnetometers with respect to the tilt of the gyros was essential. The plan for calibrating the SQUIDs was to exploit the aberration of starlight, which causes a precisely calculable misalignment between the rotors and the telescope as the latter shifts its pointing toward the guide star by up to 20 arcseconds to compensate for the orbital motion of the spacecraft and the Earth. However, three important, but unexpected, phenomena were discovered during the experiment that affected the accuracy of the results.

First, because each rotor is not exactly spherical, its principal axis rotates around its spin axis with a period of several hours, with a fixed angle between the two axes. This is the familiar "polhode" period of a spinning top and, in fact, the team used it as part of their analysis to calibrate the SQUID output. But the polhode period and angle of each rotor actually decreased monotonically with time, implying the presence of some damping mechanism, and this significantly complicated the calibration analysis. In addition, over the course of a day, each rotor was found to make occasional, seemingly random "jumps" in its orientation—some as large as 100 milliarcseconds. Some rotors displayed more frequent jumps than others. Without being able to continuously monitor the rotors' orientation, Everitt and his team couldn't fully exploit the calibrating effect of the stellar aberration in their analysis. Finally, during a planned 40-day, end-of-mission calibration phase, the team discovered that when the spacecraft was deliberately pointed away from the guide star by a large angle, the misalignment induced much larger torques on the rotors than expected. From this, they inferred that even the very small misalignments that occurred during the science phase of the mission induced torques that were probably several hundred times larger than the designers had estimated.

What ensued during the data analysis phase was worthy of a detective novel. The critical clue came from the calibration tests. Here, they took advantage of residual trapped magnetic flux on the gyroscope. (The designers used superconducting lead shielding to suppress stray fields before they cooled the niobium coated gyroscopes, but no shielding is ever perfect.) This flux adds a periodic modulation to the SQUID output, which the team used to figure out the phase and polhode angle of each rotor throughout the mission. This helped them to figure out that interactions between random patches of electrostatic potential fixed to the surface of each rotor, and similar patches on the inner surface of its spherical housing, were causing the extraneous torques. In principle, the rolling spacecraft should have suppressed these effects, but they were larger than expected. The patch interactions also accounted for the "jumps": they occurred whenever a gyro's slowly decreasing polhode period crossed an integer multiple of the spacecraft roll period. What looked like a jump of the spin direction was actually a spiraling path—known to navigators as a loxodrome. The team was able to account for all these effects in a parameterized model.

The original goal of GP-B was to measure the frame-dragging precession with an accuracy of 1%, but the problems discovered over the course of the mission dashed the initial optimism that this was possible. Although Everitt and his team were able to model the effects of the patches, they had to pay the price of the increase in error that comes from using a model with so many parameters. The experiment uncertainty quoted in the final result—roughly 20% for frame dragging—is almost totally dominated by those errors. Nevertheless, after the model was applied to each rotor, all four gyros showed consistent relativistic precessions (Fig. 1, bottom). Gyro 2 was particularly "unlucky"—it had the largest uncertainties because it suffered the most resonant jumps.

When GP-B was first conceived in the early 1960s, tests of general relativity were few and far between, and most were of limited precision. But during the ensuing decades, researchers made enormous progress in experimental gravity, performing tests of the theory by studying the solar system and binary pulsars [3]. Already by the middle 1970s, some argued that the so-called parameterized post-Newtonian (PPN) parameters that characterize metric theories of gravity, like general relativity, were already known to better accuracy than GP-B could ever achieve [4]. Given its projected high cost, critics argued for the cancellation of the GP-B mission. The counter-argument was that all such assertions involved theoretical assumptions about the class of theories encompassed by the PPN approach, and that all existing bounds on the post-Newtonian parameters involved phenomena entirely different from the precession of a gyroscope. All these issues were debated, for example, in the 1994 review of GP-B that recommended its continuation.

The most serious competition for the results from GP-B comes from the LAGEOS experiment, in which laser ranging accurately tracked the paths of two laser geodynamics satellites orbiting the earth. Relativistic frame dragging was expected to induce a small precession (around 30 milliarcseconds per year) of the orbital plane of each satellite in the direction of the Earth's rotation. However, the competing Newtonian effect of the Earth's nonspherical shape had to be subtracted to very high precision using a model of the Earth's gravity field. The first published result from LAGEOS in 1998 [5, 6] quoted an error for the frame-dragging measurement of 20 to 30%, though this result was likely too optimistic given the quality of the gravity models available at the time. Later, the GRACE geodesy mission offered dramatically improved Earth gravity models, and the analysis of the LAGEOS satellites finally yielded tests at a quoted level of approximately 10%[7].

Frame dragging has implications beyond the solar system. The incredible outpouring of energy from quasars along narrow jets of matter that stream at nearly the speed of light is most likely driven by the same frame-dragging phenomenon measured by GP-B and LAGEOS. In the case of quasars, the central body is a rapidly rotating black hole. In another example, the final inward spiral and merger of two spinning black holes involve truly wild gyrations of each body's spin axes and of the orbit, again driven by the same frame-dragging effect, and these motions are encoded in gravitational-wave signals. Laser interferometric observatories on the ground, and in the future, a similar observatory in space, may detect these gravity waves. So there is a strong link between the physics Gravity Probe B was designed to uncover and that describing some of the most energetic and cataclysmic events in the universe.

Even though it is popular lore that Einstein was right (I even wrote a book on the subject), no such book is ever completely closed in science. As we have seen with the 1998 discovery that the universe is accelerating, measuring an effect contrary to established dogma can open the door to a whole new world of understanding, as well as of mystery. The precession of a gyroscope in the gravitation field of a rotating body had never been measured before GP-B. While the results support Einstein, this didn't have to be the case. Physicists will never cease testing their basic theories, out of curiosity that new physics could exist beyond the "accepted" picture.

## About the Author

### Clifford M. Will

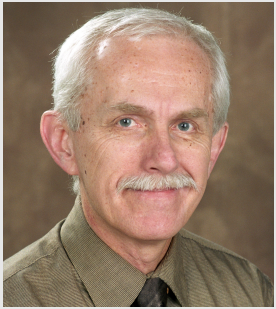

Clifford M. Will is the James S. McDonnell Professor of Space Sciences in the Physics Department at Washington University in St. Louis, and is the author of *Was Einstein Right?* From 1998 to 2011 he chaired NASA's Science Advisory Committee for Gravity Probe B.